\documentclass[twocolumn,prd,showpacs,floatfix]{revtex4}
\usepackage{epsfig}

\newcommand{\lsim}{\; ^< \!\!\!\! _\sim \;}

\begin{document}
\title{The path to metallicity: synthesis of CNO elements in standard BBN}
\author{F. Iocco$^{1,2}$, G. Mangano$^1$, G. Miele$^1$, O. Pisanti$^1$, and P.~D.~Serpico$^{3}$}

\affiliation{ $^1$Dipartimento di Scienze Fisiche Universit\`a di Napoli ``Federico II"
and INFN Sezione di Napoli, via Cintia, 80126 Naples, Italy\\
$^2$Kavli Institute for Particle Astrophysics and Cosmology
PO Box 20450, Stanford, CA 94309, USA\\
$^3$Center for Particle Astrophysics, Fermi National Accelerator
Laboratory, Batavia, IL 60510-0500 USA}

\begin{abstract}
We perform a reanalysis of the production of CNO elements in a
standard Big Bang Nucleosynthesis scenario. The CNO yields in BBN
are suppressed by the low density of the plasma, Coulomb barrier
effects and the short time scales involved. Yet, the inclusion of
nuclides and reactions traditionally disregarded may lead to an
increase relevant enough to affect the pristine Population III
stars. After a critical reanalysis and upgrade of the nuclear
network our results show no major discrepancies with the ones
obtained using a smaller nuclear network. The robustness of the
standard predictions---the early generation of star developed in a
metal-free environment---is confirmed.
\end{abstract}

\pacs{26.35.+c, 
97.20.Wt 
\hfill DSF-03/2007, FERMILAB-PUB-07-024-A, SLAC-PUB-12332}

\date{\today}
\maketitle
\section{Primordial Nucleosynthesis and Population III stars}
Standard Big Bang Nucleosynthesis (BBN) is a one-parameter theory,
mostly depending on well known nuclear and particle physics
processes at the keV-MeV scale. The only free parameter is the
baryon fraction $\omega_b$, which can be presently determined by
Cosmic Microwave Background (CMB) anisotropy~\cite{Spergel:2006hy}.
Using the CMB value for $\omega_b$ turns BBN into a parameter-free
(and thus highly predictive) theory, which can be used to check the
internal consistency of the standard cosmological model, to
constrain astrophysical mechanisms like $^7$Li and $^3$He stellar
reprocessing \cite{Eggleton:2006uc,Korn:2006tv}, to probe the cosmic
neutrino background and/or constrain exotic physics, as for example
in~\cite{Barger:2003rt,Cuoco:2003cu,Jedamzik:2004er,Cyburt:2004yc,Serpico:2005bc,Mangano:2005cc,Mangano:2006ar}.

In recent analyses
like~\cite{Cyburt:2004cq,Descouvemont:2004cw,Serpico:2004gx} the
nuclear channels for the synthesis of light elements up to mass
number $A=7$ have been carefully studied. The relevant reaction
rates have also been updated with the most recent experimental and
theoretical estimates, thus reducing the uncertainties on the light
element abundances. Indeed, a major uncertainty on nuclear
abundances arises from the uncertainty on nuclear rates. However, a
necessary pre-requisite is that all the reactions relevant for the
synthesis of the elements of interest are included. The widely used
Wagoner-Kawano code \cite{KawCode92} contains nuclides up to
${}^{16}$O, but no detailed analysis of the completeness of the
network for mass number $A>7$ is present in the literature. Indeed,
it is well known that the low density of the plasma, the higher
Coulomb barrier, and the short timescales involved in BBN conspire
to produce only small amounts of these elements. Yet, although being
of negligible entity for the ``traditional'' BBN predictions of
light element yields, missing reactions might have huge effects on
the synthesis of ``metals"---elements with mass number $A\geq12$,
whose sum we shall denote in the following with M---in particular of
the isotopes of Carbon, Nitrogen and Oxygen (CNO elements). The
astrophysical properties of stars formed from the collapse of the
very first structures in the Universe (``PopIII stars") may depend
on the chemical composition of the
cloud~\cite{Bromm:2001md,Jappsen:2005zs}. For example, metallicities
as low as 10$^{-5\pm 1}$ of the solar one (i.e. number fraction of
metals as low as M/H$\simeq 10^{-8\pm 1}$) might affect the
formation of the first generation of {\it low-mass} stars
\cite{Schneider:2003em}. These considerations motivated the present
study, whose purpose is to explore the CNO synthesis channels in BBN
and put sound constraints on their abundances by checking the
consistency of the nuclear network. We anticipate that our results
essentially confirm the robustness of the standard prediction that
the early generation of star developed in a metal-free environment.

This paper is structured as follows: in Sec. \ref{current} we
review the nucleosynthesis path leading to CNO elements in the
standard BBN code, while in Sec. \ref{update} we describe the way
this standard nuclear network has been updated and the results we
obtain. Finally, in Sec. \ref{concl} we draw our conclusions.

\section{Current nuclear network analysis}\label{current}
We can summarize the main steps of our analysis as follows:\\ i)
identification of relevant channels for CNO synthesis in the
\emph{existing} code;\\
ii) update of the relevant reaction rates which are already
present;\\
iii) addition of 4 new nuclides: ${}^{9}$Li, ${}^{10}$Be,
${}^{9}$C, ${}^{10}$C;\\
iv) addition of more than 100 nuclear reactions previously
neglected.

Here we describe the points i), ii), leaving the discussion of the
following ones to the next section. The CNO synthesis channels in
the BBN network described in \cite{Serpico:2004gx}, (hereafter
``{\sf C1}'') are basically the same as in the original
Wagoner-Kawano code (hereafter ``{\sf C0}''). One simplification
with respect to former analyses is that now one can fix the baryon
fraction from CMB measurements,
$\omega_b=0.0223_{-0.0009}^{+0.0007}$~\cite{Spergel:2006hy}, so that
within current uncertainties, the nucleosynthetic path is uniquely
determined.

The eventual small amount of Carbon is synthesized through the
direct reaction channel
\begin{equation}
^{7}{\rm Li}(\alpha,\gamma)^{11}{\rm B}(p,\gamma)^{12}{\rm C},
\end{equation}
 or the secondary one
 \begin{equation}
 ^{7}{\rm Li}(\alpha,\gamma){}^{11}{\rm B}(d,n){}^{12}{\rm C}.
 \end{equation}
None of these reactions is responsible for the final abundance of
${}^{11}$B. In fact, during BBN both these processes compete with
the stronger $^{11}{\rm B}(p,\alpha)2\:{}^4$He, which depletes all
${}^{11}$B. The final abundance of this element is therefore
provided by the $\beta$ decay of ${}^{11}$C produced by $\alpha$
capture on ${}^7$Be,
 \begin{equation}
^{7}{\rm Be}(\alpha,\gamma)^{11}{\rm C}{}\rightarrow{}^{11}{\rm B},
\end{equation}
which is Coulomb-barrier suppressed with respect to the analogous
$\alpha$ capture on ${}^7$Li.

 In principle, another possible path is via the unstable nuclei with $A=8$,
 \begin{eqnarray}
^{8}{\rm Li}(\alpha,\gamma)^{12}{\rm B}{}&\rightarrow&{}^{12}{\rm C},\\
^{8}{\rm B}(\alpha,\gamma)^{12}{\rm N}{}&\rightarrow&{}^{12}{\rm C},
\end{eqnarray}
where $^8$Li is produced e.g. by $^{7}{\rm Li}(n,\gamma)$ and
$^{6}{\rm Li}(t,p)$, and $^8$B via $^{7}{\rm Be}(p,\gamma)$.
According to our analysis, however, these channels only provide a
sub-leading contribution to the CNO abundances.

The production of heavier isotopes of Carbon proceeds typically by
radiative proton capture on carbon nuclei, followed by a $(n,p)$
reaction. For example, ${}^{13}$C is produced from light elements
through the following path
\begin{equation}
^{7}{\rm Li}(\alpha,\gamma){}^{11}{\rm B}(p,\gamma){}^{12}{\rm C}(p,\gamma){}^{13}{\rm N}(n,p){}^{13}{\rm C}
\end{equation}
and ${}^{14}$C proceeds along from ${}^{13}$C according to
\begin{equation}
{}^{13}{\rm C}(p,\gamma){}^{14}{\rm N}(n,p){}^{14}{\rm C}\:\:\:{\rm
or}\:\:\:{}^{13}{\rm C}(n,\gamma){}^{14}{\rm C}.
\end{equation}
Nitrogen is produced by means of proton radiative capture on Carbon
nuclei -- e.g. ${}^{12}{\rm C}(p,\gamma){}^{13}$N -- and Oxygen
isotopes are produced with the same mechanisms illustrated so far.
Note that once metals ($A\geq 12$) have been synthesized, being
thermodynamically more stable (higher binding energy per nucleon)
they are not disrupted back into lighter elements. This allows one
to use CNO elements as estimators of the whole metallicity M,
independently of the incompleteness of the network for $A>12$. It is
also worth noticing that the mechanism for ${}^{12}$C production in
BBN is completely different from the well known 3-$\alpha$ process
taking place in stars, as it proceeds through the formation of the
intermediate fragile nuclide ${}^{11}$B. This is easily explained in
terms of (i) the very low density of the plasma during BBN (less
than 10$^{20}$ baryons/cm$^3$ at $T\lsim 100\,$ keV); (ii) the
Coulomb suppression relevant for heavier nuclei; (iii) the short
time-scales involved, of the order of minutes.

Our analysis confirms that the synthesis of CNO elements in BBN
proceeds along a path involving intermediate mass nuclides
($4<A<12$); it is therefore conceivable that by neglecting
intermediate mass elements or ``unusual'' reactions --for stellar
astrophysics standards-- relevant channels may have been omitted.
This possibility is scrutinized in the following.

\section{Enlarging the nuclear network}\label{update}

The next steps of our analysis have been to modify the code {\sf C1}
adding 4 nuclides and more than 100 nuclear reactions previously
neglected. The criterion adopted for adding new nuclides is based on
their half-lives $t_{1/2}$. We find that 4 nuclides previously
neglected in the {\sf C0} code have a value of $t_{1/2}$ which is
comparable or longer than the typical times involved in BBN. We have
therefore included ${}^{9}$Li ($t_{1/2}$=0.178 s), ${}^{10}$Be
($t_{1/2}$=1.5$\times 10^6$ years), ${}^{9}$C ($t_{1/2}$=0.125 s),
${}^{10}$C ($t_{1/2}$=19.29 s). By adding the main reactions
involving these nuclides we are including all viable intermediate
channels connecting lighter to heavier elements. The reactions
included for these nuclides, as well as those we added for the
nuclides already included in the standard {\sf C1} version, have been
selected on the basis of physical arguments, first of all Coulomb
barrier considerations. Reactions with $Z_1Z_2 \geq 12$ have not
been considered as they are mostly suppressed in the late times of
BBN, when metals are mainly formed, and energy is very low, namely
few keV. Also, no projectiles heavier than ${}^4$He have been
considered, since Lithium and heavier element abundances are
extremely small. For each nuclide a full network including radiative
capture, stripping/pick-up and charge-exchange reactions has been
implemented. Many of these reactions were missing in the original
nuclear network of the {\sf C0}. In fact, they do not have
appreciable effects on the light nuclide abundances which was the
principal goal of the original codes. To evaluate the missing
reaction rates, extensive use has been made of compilations and
on-line libraries like~\cite{audi97,toi20,endf}. The nuclear rates
for many of these reactions, when measurements or theoretical models
were missing, have been estimated with simple nuclear models, of the
kind presented in~\cite{FH64,wfh67}. An order of magnitude
uncertainty has been assumed, according to Wagoner's prescription.
We shall refer to the code including the new network {\it and} the
new nuclides as ``{\sf C3}'' code (30 nuclides and 262 reactions),
while the code with an enlarged network but not including the
four above-mentioned nuclides is named ``{\sf C2}'' (26 nuclides and
240 reactions). The {\sf C2,3} codes show only few additional
subleading synthesis channels for CNO elements with respect to the
{\sf C0,1} codes, in particular a minor difference in ${}^{12}{\rm
C}$ (due to the additional subleading channel $^{11}{\rm
B}(d,p){}^{12}{\rm B}\to{}^{12}{\rm C}$), and a slightly larger
amount of ${}^{13}{\rm C}$, ${}^{14}{\rm C}$ and ${}^{14}{\rm N}$,
see Table~\ref{tab:tab1}. The increase of ${}^{13}{\rm C}$ and
${}^{14}{\rm C}$ is due to a more efficient burning of ${}^{12}{\rm
C}$ through the new channels $^{12}{\rm C}(d,p){}^{13}{\rm C}$ and
$^{13}{\rm C}(d,p){}^{14}{\rm C}$. Due to the increasing of
${}^{13}{\rm C}$, ${}^{14}{\rm N}$ abundance is also increased by
the (already present) reaction $^{13}{\rm C}(p,\gamma){}^{14}{\rm
N}$.

\begin{figure}[!htbp]
\begin{center}
\epsfig{file=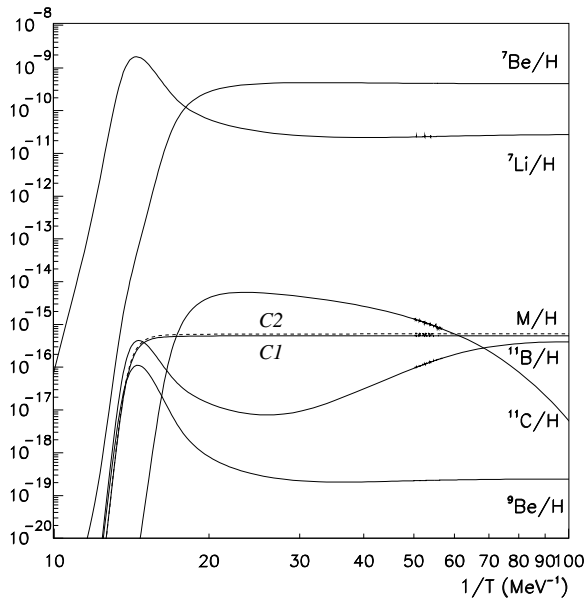,width=8.0truecm} \caption{The metal abundance
M/H in the {\sf C1} (solid line) and {\sf C2} (dashed line) codes,
compared with the abundances of a few lighter
elements.}\label{BBNevol}
\end{center}
\end{figure}

On the other hand we find that the {\sf C2} and {\sf C3} codes show
no appreciable difference in the final abundances, thus proving that
the 4 added nuclides play no role in CNO synthesis, as reported in
Table~\ref{tab:tab1}. There are two clear conclusions that one can
draw from our results: (i) an accurate (i.e. at the 10\% level)
prediction of the metal yield in standard BBN {\it does depend} on
the details of the intermediate mass nuclide reaction network; (ii)
no {\it major} change is found when enlarging the network. The
physical reason is that at the beginning of the BBN, when the higher
temperature and density would favor the formation of CNO elements,
their synthesis is suppressed in a high entropy Universe and by the
lack of necessary intermediate elements. On the other hand, when
enough ``catalyzers'' are produced at later times both temperature
and density are not high enough to overcome the high Coulomb
barrier. The evolution of metals and a few lighter elements vs.
temperature is shown in Fig.~\ref{BBNevol}. Note the correlation of
the $^{9}{\rm Be},\,{}^{11}{\rm B},\,$ and metal yields with the
peak in the ${}^{7}{\rm Li}$ production. Also note that the metal
yields have a monotone behavior, due to their higher stability with
respect to lighter elements.

Even taking into account a much enlarged network, one might still
wonder how robust are the previous conclusions, especially given the
lack of data or detailed theoretical models for several reactions. A
detailed discussion of uncertainties for the order 100 new reactions
is clearly unpractical. To obtain a generous but more robust upper
limit to the yield of metals in standard BBN, we have performed an
analysis of their yields assuming that all the reactions producing
at some stage an element with $A> 7$ ``instantaneously'' produce
${}^{12}$C, instead. In other words, a strong upper bound is
obtained by assuming that all yields for nuclides with $A>7$
contribute to CNO eventually. In this way, we neglect any
destruction mechanism of intermediate nuclei, the multistep nature
of the synthesis of CNO nuclides, and we are maximizing the time
available for their synthesis. As a result, we get ${\rm M/H}\alt
10^{-10}$. This bound is clearly very conservative (about a factor
10$^5$ higher than our best estimate), especially considering the
fact that nuclides with $7<A<12$ are very fragile, and the leading
reactions they are involved in typically tend to destroy them rather
than producing CNO elements. It is also quite robust, as it
basically depends upon the thermodynamical properties of the plasma
holding for standard BBN and on the well-known nuclear rates
involving target nuclei with $A\leq 7$ only.

\begin{widetext}
\begin{center}
\begin{table}[!htb]
\begin{tabular}{|c||c|c|c|c|c|c|c|}
  \hline
  Nuclide &
  ${}^{9}$Be/H &
  ${}^{11}$B/H &
  ${}^{12}$C/H &
 ${}^{13}$C/H &
 ${}^{14}$C/H &
 ${}^{14}$N/H &
 ${}^{16}$O/H \\
   &
  $(\times 10^{-19}$)&
  $(\times 10^{-16}$)&
  $(\times 10^{-16}$)&
  $(\times 10^{-17}$)&
  $(\times 10^{-17}$)&
  $(\times 10^{-17}$)&
  $(\times 10^{-20}$) \\
\hline\hline
  {\sf C1} &
 2.4 &
 3.9 &
 4.4 &
 7.6&
 0.6 &
 2.6 &
 1.8 \\
\hline
  {\sf C2} &
  2.5 &
  3.9 &
  4.6 &
 9.0 &
 1.3 &
 3.7 &
 2.7 \\
  \hline
  {\sf C3} &
  2.5 &
  3.9 &
  4.6 &
 9.0 &
 1.3 &
 3.7 &
 2.7 \\
  \hline
\end{tabular}
\caption{\label{tab:tab1} Some "heavier" nuclei yields (for
$\omega_b$=0.0223) predicted by the {\sf C1} code and by using the
upper limits to the production rates in the {\sf C2} and {\sf C3}
codes, described in the text.}
\end{table}
\end{center}
\end{widetext}

\section{Conclusions}\label{concl}
Besides predicting detailed values for the abundances of the light
nuclei $^2$H, $^3$He, $^4$He, and  $^7$Li, standard BBN also
predicts that the first collapsed objects in the Universe should
form in a metal free gas. In fact, Carbon is produced along inefficient
paths involving intermediate mass elements, in particular ${}^{11}$B
rather than the usual $3\alpha$ reaction in stars. Very small traces
of Nitrogen and Oxygen are then produced by radiative capture upon
${}^{12}$C. These predictions are relevant for determining the
physical mechanisms regulating the collapse of the clouds leading to
the PopIII stars, and the evolution of the smallest among these
pristine stars.

Given the relevance of this topic, and the incompleteness of
standard BBN nuclear networks in the mass range $A>$7, we have
performed a detailed study of the synthesis channels of Carbon,
Nitrogen and Oxygen in Primordial Nucleosynthesis. We have added to
the standard code 4 nuclides and more than 100 reactions observing
no sensible increase of CNO elements. Re-analysing the synthesis of
CNO after the addition of the new reactions we find that the main
channels for their production are basically the same as before and
none of the newly added reactions/nuclides opens effective channels
from light to heavy elements in BBN. We consider this as a robust
check that only negligible traces of metals are produced in {\it
standard} BBN, in agreement with earlier and less accurate analyses.
This should be regarded as a further observational test for standard
BBN, since alternative theories like Inhomogeneous BBN might lead to
primordial metallicities even larger than $10^{-6}$ of the solar
value (see e.g.~\cite{Jedamzik:1994de,Jedamzik:1999jb}).

\section*{Acknowledgments}
We thank T. Abel and R. Wagoner for useful discussions. In Naples,
this work was supported in part by PRIN04 of the Italian MIUR under
grant Fisica Astroparticellare. PDS acknowledges support by the US
Department of Energy and by NASA grant NAG 5-10842 at Fermilab.

\end{document}